\documentclass[aps,floatfix,showpacs,nofootinbib,showkeys,superscriptaddress,notitlepage]{revtex4}
\usepackage[T1]{fontenc}
\usepackage[english]{babel}
\usepackage[dvips]{graphicx}
\usepackage{amsmath,amssymb,amsfonts,mathtext,enumerate,float,mathrsfs}
\usepackage{wrapfig}
\usepackage{hhline}

\usepackage{epsfig}
\usepackage{epsf}
\usepackage{subfig} 
\usepackage{feynmp} 
\usepackage{pgf,pgfarrows,pgfnodes,pgfautomata,pgfheaps,multirow}
\usepackage{array}
\usepackage{bm}
\usepackage{latexsym}
\usepackage{pifont}

\begin{document}
\title{Radiative decays of radially excited pseudoscalar mesons in the extended Nambu--Jona-Lasinio model}

\author{A.\ V.\ Vishneva}
\email{vishneva@theor.jinr.ru}
\affiliation{Bogoliubov Laboratory of Theoretical Physics,
JINR, Dubna, 141980  Russia}

\author{M.\ K.\ Volkov}
\email{volkov@theor.jinr.ru}
\affiliation{Bogoliubov Laboratory of Theoretical Physics,
JINR, Dubna, 141980  Russia}

\begin{abstract}

In the extended NJL model the radiative decay widths of the radially excited states of the pseudoscalar $\pi$, $\eta$, and $\eta'$ mesons are calculated. The predictions for the decay widths of the processes $\pi(1300)\longrightarrow(\rho^0,\omega)\gamma$, $\eta(1295)\longrightarrow(\rho^0,\omega,\phi)\gamma$, and $\eta(1475)\longrightarrow(\rho^0,\omega,\phi)\gamma$ are given. Nowadays, there are no solid experimental data for these processes. The comparison of the results obtained in the framework of the standard and the extended NJL models for decays of the ground states of mesons is given. It is shown that these calculations correspond to each other and are also in satisfactory agreement with experimental data. This allows one to expect that the extended NJL can give reliable predictions for the excited states of mesons.
\end{abstract}



\maketitle

\section*{INTRODUCTION}

Taking into account the excited states of mesons plays an important role in the description of hadronic interactions at low energies. Nowadays, the properties of the radially excited mesons are studied at different colliders like VEPP-2000 (Novosibirsk), BEPC-II (Beijing) etc.
 
However, we still do not have any solid experimental data in this domain. Therefore, the theoretical study of the excited states of mesons is of apparent interest. The present work is devoted to the solution of one of these problems.

Let us note that for description of the low-energy interactions of mesons in the ground states the $U(3)\times U(3)$ standard Nambu--Jona-Lasinio (NJL) model can be successfully used \cite{v86,er,vw,kl,v93,ver,ufn}.
On the other hand, for the description of the first radial excitations of mesons the extended NJL model was proposed \cite{ufn,weiss,yaf,yud,ven}.
In this model, the excited states are described with a form factor in the form
 $f_a(\textbf{k})=c_a(1+d_a \textbf{k}^2)$, where $d_a$ is a slope parameter, \textbf{k} is transversal quark momentum, and $c_a$ is a constant defining the meson masses. As a result, the mass spectrum and strong, weak, and electromagnetic interactions of mesons were described \cite{yud, ven,akv1,akv2,arb,eepipi,taupipi,eta}.

Let us emphasize that extra difficulties appear in the description of the $\eta$ and $\eta'$ mesons. Indeed, even considering only ground states one has to take into account the singlet-octet mixing caused by the gluon anomaly ($U_A (1)$-problem). This problem is usually solved by adding the 't Hooft interaction \cite{vw,kl,ver}.

However, considering radially excited states of $\eta$ and $\eta'$ mesons one has to take into account the mixing of the ground and excited states. In order to describe both mixings simultaneously, it is necessary to use the $4\times4$ matrix. This matrix was obtained in \cite{matrix} and successfully applied to describe the masses of $\eta$ and $\eta'$ mesons and the processes involving them \cite{yud,arb,eta}.
In the present work this matrix is used for the description of radiative decays of the radially excited $\eta$ and $\eta'$ mesons.

The article is organized as follows. In the next section we introduce the Lagrangians of the $U(3)\times U(3)$ standard and the extended NJL models describing the quark-meson interactions. In section 3, the amplitudes and widths of the radiative decays of the processes considered in this paper are given. The comparison of calculations within the standard and extended NJL models with experimental data is also shown. The predictions for the decay widths of the excited pseudoscalar mesons are made in section 4. In conclusion our results are briefly discussed. In the Appendix the problem concerning pseudoscalar--axial-vector (P-A) transitions for $\eta$ and $\eta'$ is discussed. We give the alternative results for the processes considered here without taking into account P-A transitions.

\section{Lagrangians of the standard and the extended NJL models}

The processes considered in this paper are shown on Fig.~1.

\begin{figure}[!h]
\includegraphics[width=0.45\linewidth]{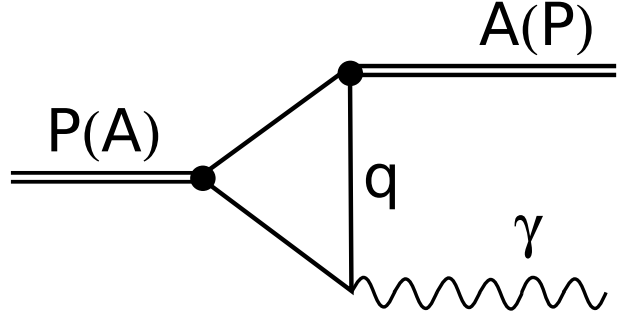}
\caption{Quark triangle diagram describing radiative dacays of pseudoscalar (P) and vector (V) mesons. $q=(u,d,s)$ are quark fields.}
\label{fig1}
\end{figure}

The corresponding Lagrangian of the extended NJL reads \cite{arb,eta}

\begin{equation}
\mathcal L = \bar{q}(k')\left( L_{\mathrm{f}} + L_\gamma + L_\mathrm{V}(p) + L_{P}(p)\right) q(k)\,,
\end{equation}

where $p=k-k'$,

$L_{\mathrm{f}} = i\hat{\partial} - m \,,$

$L_\gamma = \frac{e}{2}\left(\lambda_3+\frac{\lambda_8}{\sqrt{3}}\right)\hat{A}\,,$

$L_\mathrm{V}(p) = A_{\omega,\rho}\left(\lambda_3{\hat{\rho}}(p)+\lambda_u\hat{\omega}(p)\right) + A_\phi \lambda_s\hat{\phi}(p)- A_{\omega',\rho'}\left(\lambda_3{\hat\rho'}(p)+\lambda_u\hat\omega'(p)\right) - A_{\phi'} \lambda_s\hat{\phi'}(p)\,,$

$L_{P}(p) = \left(i\gamma_5\sum\limits_{q=u,s}\lambda_q\sum\limits_{\boldsymbol\eta = \eta,\eta',\hat\eta,\hat\eta'}A_{\boldsymbol{\eta}}^q\eta(p)\right) + A_{\pi} \lambda_3 \gamma_5 \pi^0 (p) -A_{\pi'} \lambda_3 \gamma_5 \pi'(p),$ 

Here $\bar{q} = (\bar{u},\bar{d},\bar{s})$ are quark fields; $m = \mathrm{diag}(m_u,m_d,m_s)$, where $m_u = m_d = $ 280 MeV, $m_s =$ 405 MeV; $\hat{A}$ is the photon field; $\rho(\rho')$, $\omega(\omega')$ and $\phi(\phi')$ are the vector meson fields and $\pi(\pi')$, $\eta(\hat{\eta})$ and $\eta'(\hat{\eta'})$ are the pseudoscalar meson fields in the ground and excited states, respectively; $\lambda_3$ and $\lambda_8$ are the Gell-Mann matrices, $\lambda_u=\mathrm{diag}(1,1,0), \lambda_s=\mathrm{diag}(0,0,-\sqrt{2})$;

\begin{eqnarray}
A_{\omega,\rho} &=& g_{\rho_1}\frac{\sin(\beta^u+\beta_0^u)}{\sin(2\beta_0^u)}
       +g_{\rho_2}f_u({k^\bot}^2)\frac{\sin(\beta^u-\beta_0^u)}{\sin(2\beta_0^u)},
\\
A_{\omega',\rho'} &=& g_{\rho_1}\frac{\cos(\beta^u+\beta_0^u)}{\sin(2\beta_0^u)}
        +g_{\rho_2}f_u({k^\bot}^2)\frac{\cos(\beta^u-\beta_0^u)}{\sin(2\beta_0^u)},
\nonumber \\
A_{\phi} &=& g_{\phi_1}\frac{\sin(\beta^s+\beta_0^s)}{\sin(2\beta_0^s)}
       +g_{\phi_2}f_s({k^\bot}^2)\frac{\sin(\beta^s-\beta_0^s)}{\sin(2\beta_0^s)},
\nonumber \\
A_{\phi'} &=& g_{\phi_1}\frac{\cos(\beta^s+\beta_0^s)}{\sin(2\beta_0^s)}
        +g_{\phi_2}f_s({k^\bot}^2)\frac{\cos(\beta^s-\beta_0^s)}{\sin(2\beta_0^s)}, \nonumber \\
A_{\pi} &=& g_{\pi_1}\frac{\sin(\alpha+\alpha_0)}{\sin(2\alpha_0)}
       +g_{\pi_2}f_s({k^\bot}^2)\frac{\sin(\alpha-\alpha_0)}{\sin(2\alpha_0)},
\nonumber \\
A_{\pi'} &=& g_{\pi_1}\frac{\cos(\alpha+\alpha_0)}{\sin(2\alpha_0)}
        +g_{\pi_2}f_s({k^\bot}^2)\frac{\cos(\alpha-\alpha_0)}{\sin(2\alpha_0)}, \nonumber
\end{eqnarray}
the mixing angles are $\alpha_0=59.06^{\circ}, \alpha=59.38^{\circ}, \beta_0^u=61.44^{\circ}, \beta^u=79.85^{\circ}, \beta_0^s=57.11^{\circ}, \beta^s=76.18^{\circ}$; the slope parameters in the form-factors are $d_u=-1.78~ \mathrm{GeV}^{-2}$, $d_s=-1.73~\mathrm{GeV}^{-2}$;

\begin{equation}
A_{\eta,\hat\eta,\eta',\hat\eta'}^q = g_{q_1}b_{\eta,\hat\eta,\eta',\hat\eta'}^{\varphi_{q,1}} + g_{q_2}b_{\eta,\hat\eta,\eta',\hat\eta'}^{\varphi_{q,2}}f_u({k^\bot}^2),
\end{equation}
where q=u,s; the mixing coefficients $b_{\eta,\hat\eta,\eta',\hat\eta'}^{\varphi_{q,i}}$ are given in Table I.

\begin{table}
\caption{The mixing coefficients $b_{\eta,\hat\eta,\eta',\hat\eta'}^{\varphi_{q,i}}$ for the isoscalar pseudoscalar meson states.}
\begin{ruledtabular}
\begin{tabular}{rcccc}	 		
&$\eta$ 		&$\hat\eta$ 	&$\eta'$ 		&$\hat\eta'$\\
$\varphi_{u,1}$	&$0.71$		&$0.62$		&$-0.32$		&$0.56$		\\
$\varphi_{u,2}$   &$0.11$		&$-0.87$		&$-0.48$		&$-0.54$		\\
$\varphi_{s,1}$	&$0.62$		&$0.19$		&$0.56$		&$-0.67$		\\
$\varphi_{s,2}$   &$0.06$		&$-0.66$		&$0.30$		&$0.82$		\\
\end{tabular}
\end{ruledtabular}
\label{table:1}
\end{table}

The coupling constants $g_{\pi_i}=g_{u_i}$, $g_{s_i}$, $g_{\rho_i} $ and $g_{\phi_i}$, where i=1 for the ground state and i=2 for the first radial excitations, have the form

\begin{eqnarray}
g_{q_1} = \left(4 \frac{I_2(m_q)}{Z_q}\right)^{-1/2},\quad g_{q_2} = \left(4 I_2^{f^2}(m_q)\right)^{-1/2}, \\
g_{\rho_1} = \left(\frac{2}{3} I_2(m_u)\right)^{-1/2},\quad g_{\rho_2} = \left(\frac{2}{3} I_2^{f^2}(m_u)\right)^{-1/2}, \nonumber \\
g_{\phi_1} = \left(\frac{2}{3} I_2(m_s)\right)^{-1/2},\quad g_{\phi_2} = \left(\frac{2}{3} I_2^{f^2}(m_s)\right)^{-1/2}, \nonumber
\end{eqnarray}

the factor $Z_q$ appears from P-A transitions, $Z_u\simeq Z_s = 1.2$ (see also Appendix),

\begin{equation}
I^{f^{n}}_m(m_q) = \int\frac{\mbox{d}^4 k}{(2\pi)^4}\frac{(f_q({k^\bot}^2))^n}{(m_q^2-k^2)^m}\Theta(\Lambda^2_3 - \vec k^2),
\end{equation}
the cut-off parameter $\Lambda^2_3=1.03$ GeV.

In the framework of the standard NJL the Lagrangian has a simpler form:
\cite{v86,arb}:

\begin{equation}
\mathcal L = \bar{q}\left( L_{\mathrm{f}} + L_\gamma + L_\mathrm{V} + L_{P}\right) q\,,
\end{equation}
where

$L_{\mathrm{f}} = i\hat{\partial} - m \,,$

$L_\gamma = \frac{e}{2}\left(\lambda_3+\frac{\lambda_8}{\sqrt{3}}\right)\hat{A}\,,$

$L_\mathrm{V} = \frac{g_{\rho_1}}{2}\left(\lambda_3{\hat{\rho}}+\lambda_u\hat{\omega}\right) + \frac{g_{\phi_1}}{2} \lambda_s\hat{\phi}\,,$

$L_\mathrm{P} = i g_{\pi} \gamma_5 \lambda_3 \pi^0 + i \gamma_5 \eta \left( \lambda_u g_{\pi} \sin \bar{\theta}+\lambda_s g_{s} \cos \bar{\theta}\right)+i \gamma_5 \eta' \left( \lambda_u g_{\pi} \cos \bar{\theta}-\lambda_s g_{s} \sin \bar{\theta}\right),$
where $\bar{\theta}=54^{\circ}$, and the coupling constants are $g_\pi=m_u/F_\pi$, $g_s=m_s/F_s$, $F_\pi=93~\mbox{MeV}$, $F_s=1.28 F_\pi$.

\section{Decay widths for the ground states of the mesons in the standard and the extended NJL}

The widths of the decays shown on Fig.1 in the framework of the standard NJL take the form

\begin{eqnarray}
\Gamma_{NJL}(V\rightarrow P\gamma)= \frac{\alpha \alpha_\rho}{6} \frac{C^2_{VP}}{F^2_P} \left(\frac{M_V^2-M_P^2}{4\pi M_V^2}\right)^3, \\
\Gamma_{NJL}(P\rightarrow V\gamma)= 3 \frac{\alpha \alpha_\rho}{6} \frac{C^2_{VP}}{F^2_P} \left(\frac{M_P^2-M_V^2}{4\pi M_P^2}\right)^3,
\end{eqnarray}

where $\alpha_\rho=g_\rho^2/4\pi$,

\begin{eqnarray}
C_{\rho\pi}=1, \quad C_{\omega\pi}=3, C_{\omega\eta}=\sin\bar{\theta}, \quad C_{\rho\eta}=3 \sin \bar{\theta}, \\
C_{\phi\eta}=2 \cos\bar{\theta}, \quad C_{\rho\eta'}=3 \cos \bar{\theta}, \quad C_{\omega\eta'}=\cos\bar{\theta}, \quad C_{\phi\eta'}=2 \sin\bar{\theta}. \nonumber
\end{eqnarray}

In the extended NJL the decay widths take the form

\begin{eqnarray}
\Gamma_{ENJL}(V\rightarrow P\gamma)= \frac{6\alpha}{(16 \pi^2 m_q)^2} V_{VP}^2 \left(\frac{M_V^2-M_P^2}{M_V^2}\right)^3, \\
\Gamma_{ENJL}(P\rightarrow V\gamma)= 3 \frac{6\alpha}{(16 \pi^2 m_q)^2} V_{VP}^2 \left(\frac{M_P^2-M_V^2}{M_P^2}\right)^3,
\end{eqnarray}

where

\begin{eqnarray}
V_{V\pi} = g_{V_1}\left(\frac{\sin(\beta^u+\beta_0^u)} {\sin(2\beta_0^u)} g_{\rho_1}I_3(m_u)+\frac{\sin(\beta^u-\beta_0^u)} {\sin(2\beta_0^u)} g_{\rho_2}I_3(m_u)\right),\\
V_{V\eta_i} = \frac{\sin(\beta^{q}+\beta_0^{q})}{\sin(2\beta_0^{q})}b_{\eta,\eta'}^{\varphi_{q,1}}g_{V_1}g_{q_1} I_3(m_q) +
\frac{\sin(\beta^{q}-\beta_0^{q})}{\sin(2\beta_0^{q})}b_{\eta,\eta'}^{\varphi_{q,1}}g_{V_2}g_{q_1} I^{f}_3(m_q) + \nonumber \\
\frac{\sin(\beta^{q}+\beta_0^{q})}{\sin(2\beta_0^{q})}b_{\eta,\eta'}^{\varphi_{q,2}}g_{V_1}g_{q_2} I^{f}_3(m_q) +
\frac{\sin(\beta^{q}-\beta_0^{q})}{\sin(2\beta_0^{q})}b_{\eta,\eta'}^{\varphi_{q,2}}g_{V_2}g_{q_2} I^{f^2}_3(m_q),
\end{eqnarray}
where V is $\rho, \omega,\phi$ and $\eta_i$ are $\eta,\eta'$; $q=u$ for $\rho$ and $\omega$ and $q=s$ for $\phi$.
Using these expressions we obtained the widths of the decays of the ground states of the mesons given in Table II.
 
\begin{table}
\caption{The decay widths of the ground states of the mesons within the standard and the extended NJL}
\begin{ruledtabular}
\begin{tabular}{|c|c|c|c|}
\hline 
\rule[-1ex]{0pt}{2.5ex} Decay & Standard NJL (keV) & Extended NJL (keV) & PDG (keV) \cite{pdg} \\ 
\hline 
\rule[-1ex]{0pt}{2.5ex} $\rho\rightarrow\pi\gamma$ & 89.8 & 81.7 & 89.4 \\ 
\hline 
\rule[-1ex]{0pt}{2.5ex} $\omega\rightarrow\pi\gamma$ & 832 & 756 & 703 \\
\hline 
\rule[-1ex]{0pt}{2.5ex} $\rho\rightarrow\eta\gamma$ & 72.9 & 63.4 & 44.7 \\ 
\hline 
\rule[-1ex]{0pt}{2.5ex} $\omega\rightarrow\eta\gamma$ & 8.7 & 7.6 & 3.9 \\ 
\hline 
\rule[-1ex]{0pt}{2.5ex} $\phi\rightarrow\eta\gamma$ & 54.6 & 51.8 & 55.3 \\ 
\hline 
\rule[-1ex]{0pt}{2.5ex} $\phi\rightarrow\eta'\gamma$ & 0.47 & 0.35 & 0.27 \\ 
\hline 
\rule[-1ex]{0pt}{2.5ex} $\eta'\rightarrow\rho\gamma$ & 69.9 & 82.3 & 57.7 \\ 
\hline 
\rule[-1ex]{0pt}{2.5ex} $\eta'\rightarrow\omega\gamma$ & 7.05 & 8.2 & 5.5 \\ 
\hline 
\end{tabular} 
\end{ruledtabular}
\label{table:2}
\end{table}

Let us note that the values presented in the second column of Table II agree with the previous ones given in \cite{ver,v93}. It is interesting to emphasize that the results obtained within the standard and the extended NJL model are close to each other and to the experiment. Thus, we suppose that the extended NJL model can give trustworthy predictions for the decay widths of the excited mesons.

\section{The decay widths of radially excited pseudoscalar mesons}

The coefficients $V_{VP}$ for amplitudes describing these decays read

 \begin{eqnarray}
V_{V{\pi'}} = -\left(\frac{\sin(\beta^{q}+\beta_0^{q})}{\sin(2\beta_0^{q})}\frac{\cos(\alpha+\alpha_0)}{\sin(2\alpha_0)}g_{V_1}g_{q_1} I_3(m_q) +
\frac{\sin(\beta^{q}-\beta_0^{q})}{\sin(2\beta_0^{q})}\frac{\cos(\alpha+\alpha_0)}{\sin(2\alpha_0)}g_{V_2}g_{q_1} I^{f}_3(m_q) +\right. \nonumber \\
\left. + \frac{\sin(\beta^{q}+\beta_0^{q})}{\sin(2\beta_0^{q})}\frac{\cos(\alpha-\alpha_0)}{\sin(2\alpha_0)}g_{V_1}g_{q_2} I^{f}_3(m_q) +
\frac{\sin(\beta^{q}-\beta_0^{q})}{\sin(2\beta_0^{q})}\frac{\cos(\alpha-\alpha_0)}{\sin(2\alpha_0)}g_{V_2}g_{q_2} I^{f^2}_3(m_q)\right),\\
V_{V\hat{\eta_i}} = \frac{\sin(\beta^{q}+\beta_0^{q})}{\sin(2\beta_0^{q})}b_{\hat{\eta},\hat{\eta'}}^{\varphi_{q,1}}g_{V_1}g_{q_1} I_3(m_q) +
\frac{\sin(\beta^{q}-\beta_0^{q})}{\sin(2\beta_0^{q})}b_{\hat{\eta},\hat{\eta'}}^{\varphi_{q,1}}g_{V_2}g_{q_1} I^{f}_3(m_q) + \nonumber \\
+ \frac{\sin(\beta^{q}+\beta_0^{q})}{\sin(2\beta_0^{q})}b_{\hat{\eta},\hat{\eta'}}^{\varphi_{q,2}}g_{V_1}g_{q_2} I^{f}_3(m_q) +
\frac{\sin(\beta^{q}-\beta_0^{q})}{\sin(2\beta_0^{q})}b_{\hat{\eta},\hat{\eta'}}^{\varphi_{q,2}}g_{V_2}g_{q_2} I^{f^2}_3(m_q),
\end{eqnarray}

where $\hat{\eta_i}$ are $\hat{\eta},\hat{\eta'}$.
The corresponding results are shown in Table III.

\begin{table}
\caption{The predictions for the decay widths of the excited mesons within the extended NJL}
\begin{ruledtabular}
\begin{tabular}{|c|c|}
\hline 
\rule[-1ex]{0pt}{2.5ex} Decay & Extended NJ, (keV) \\ 
\hline 
\rule[-1ex]{0pt}{2.5ex} $\pi(1300)\rightarrow\rho\gamma$ & 26.2  \\ 
\hline 
\rule[-1ex]{0pt}{2.5ex} $\pi(1300)\rightarrow\omega\gamma$ & 229  \\
\hline 
\rule[-1ex]{0pt}{2.5ex} $\eta(1295)\rightarrow\rho\gamma$ & 0.057  \\ 
\hline 
\rule[-1ex]{0pt}{2.5ex} $\eta(1295)\rightarrow\omega\gamma$ & 0.006  \\ 
\hline 
\rule[-1ex]{0pt}{2.5ex} $\eta(1295)\rightarrow\phi\gamma$ & 6.6  \\ 
\hline 
\rule[-1ex]{0pt}{2.5ex} $\eta(1475)\rightarrow\rho\gamma$ & 61.4  \\ 
\hline 
\rule[-1ex]{0pt}{2.5ex} $\eta(1475)\rightarrow\omega\gamma$ & 6.7  \\ 
\hline 
\rule[-1ex]{0pt}{2.5ex} $\eta(1475)\rightarrow\phi\gamma$ & 2.6  \\ 
\hline 
\end{tabular} 
\end{ruledtabular}
\label{table:3}
\end{table}

\section*{CONCLUSION}

Our calculations for the ground states of mesons show that the results obtained within the standard and the extended NJL model correspond to each other. One can see that the agreement between the theoretical and experimental results for $\eta$ and $\eta'$ is not as good as for pions. It can be explained by the fact that in the case of pions the chiral symmetry breaking is connected only with small non-zero current quark mass, and in the case of $\eta$ and $\eta'$ mesons there is also heavier s-quark and singlet-octet mixing. Therefore, in this case we have stronger chiral symmetry breaking.

Unfortunately, there are no any reliable experimental data to test our calculations for the excited mesons. However, we hope that our predictions will be verified soon.

\section*{ACKNOWLEDGEMENT}

We are grateful to A.B. Arbuzov and D.G. Kostunin for useful discussions.

\section*{APPENDIX: PROBLEMS CONCERNING P-A TRANSITIONS.}

Let us remind that taking P-A transition into account noticeably changes the coupling constants of the pseudoscalar mesons \cite{v86,ver}. However, the case of pions and kaons differs from that of $\eta$ and $\eta'$ mesons. Indeed, pions and kaons have the direct axial-vector partners. This leads to transition of the form $\pi\leftrightarrow a_1, K\leftrightarrow K_1$. However, for $\eta$ and $\eta'$ mesons the corresponding partners among the isoscalar axial-vector mesons do not exist. Therefore, in the case of $\eta$ and $\eta'$ mesons additional renormalization is possibly absent or has a more complicated form. Here we present the results without taking P-A transitions into account. In this case, we simply neglect the factor $Z$ in Eq.(4).

The corresponding decay widths for the processes involving isoscalar pseudoscalar mesons are given in Table IV.

\begin{table}
\caption{The decay widths without taking into account of P-A transitions }
\begin{ruledtabular}
\begin{tabular}{|c|c|c|c|}
\hline 
\rule[-1ex]{0pt}{2.5ex} Decay & Standard NJL (keV) & Extended NJL (keV) & PDG (keV) \cite{pdg} \\ 
\hline 
\rule[-1ex]{0pt}{2.5ex} $\rho\rightarrow\eta\gamma$ & 52.1 & 45.5 & 44.7 \\ 
\hline 
\rule[-1ex]{0pt}{2.5ex} $\omega\rightarrow\eta\gamma$ & 6.2 & 5.5 & 3.9 \\ 
\hline 
\rule[-1ex]{0pt}{2.5ex} $\phi\rightarrow\eta\gamma$ & 30.9 & 32.01 & 55.3 \\ 
\hline 
\rule[-1ex]{0pt}{2.5ex} $\phi\rightarrow\eta'\gamma$ & 0.267 & 0.257 & 0.27 \\ 
\hline 
\rule[-1ex]{0pt}{2.5ex} $\eta'\rightarrow\rho\gamma$ & 49.9 & 69.4 & 57.7 \\ 
\hline 
\rule[-1ex]{0pt}{2.5ex} $\eta'\rightarrow\omega\gamma$ & 5.03 & 6.9 & 5.5 \\ 
\hline 
\rule[-1ex]{0pt}{2.5ex} $\eta(1295)\rightarrow\rho\gamma$ & --- & 12.6 & --- \\ 
\hline 
\rule[-1ex]{0pt}{2.5ex} $\eta(1295)\rightarrow\omega\gamma$ & --- & 1.4 & --- \\ 
\hline 
\rule[-1ex]{0pt}{2.5ex} $\eta(1295)\rightarrow\phi\gamma$ & --- & 7.6 & --- \\ 
\hline 
\rule[-1ex]{0pt}{2.5ex} $\eta(1475)\rightarrow\rho\gamma$ & --- & 11.9 & --- \\ 
\hline 
\rule[-1ex]{0pt}{2.5ex} $\eta(1475)\rightarrow\omega\gamma$ & --- & 2.9 & --- \\ 
\hline 
\rule[-1ex]{0pt}{2.5ex} $\eta(1475)\rightarrow\phi\gamma$ & --- & 8.1 & --- \\ 
\hline 
\end{tabular}
\end{ruledtabular}
\label{table:4}
\end{table}

One can see that in this case there is better agreement with experiment. Therefore, the corresponding predictions for the processes with excited states can be more reliable.


\begin{thebibliography}{00}
\bibitem{v86} \textit{Volkov M.K.} Low-Energy Meson Physics in the Quark Model of Superconductivity Type (in Russian) // Sov. J. Part. and Nuclei. 1986. V. 17. P. 186.
 
\bibitem{er} \textit{Ebert D., Reinhardt H.} Effective Chiral Hadron Lagrangian with Anomalies and Skyrme Terms from Quark Flavor Dynamics // Nucl. Phys. B. 1986. V. 271. P. 188.

\bibitem{vw} \textit{Vogl U., Weise W.} The Nambu and Jona-Lasonio Model: Its Implications for Hadrons and Nuclei // Prog. Part. Nucl. Phys. 1991. V. 27. P. 195.

\bibitem{kl} \textit{Klevansky S.P.} The Nambu--Jona-Lasinio Model of Quantum Chromodynamics // Rev. Mod. Phys. 1992. V. 64. No. 3. P. 649.

\bibitem{v93} \textit{Volkov M.K.} Effective Chiral Lagrangians and Nambu--Jona-Lasinio Model // Sov. J. Part. and Nuclei. 1993. V. 24. P. 81.

\bibitem{ver} \textit{Ebert D., Reinhardt H., Volkov M.K.} Effective Hadron Theory of QCD // Prog. Part. Nucl. Phys. 1994. V. 33. P. 1.

\bibitem{ufn} \textit{Volkov M.K., Radzhabov A.E.} The Nambu--Jona-Lasinio Model and Its Development // Phys. Usp. 2006. V. 49. P. 551.

\bibitem{weiss} \textit{Volkov M.K., Weiss C.} A Chiral Lagrangian for Excited Pions // Phys. Rev. D. 1997. V. 56. P. 221-229.

\bibitem{yaf} \textit{Volkov M.K.} Excited Pseudoscalar and Vector Mesons in the $U(3)\times U(3)$ Chiral Model // Phys. At. Nucl. 1997. V. 60. No. 11. P. 2094-2103.

\bibitem{yud} \textit{Volkov M.K., Yudichev V.L.} Radially Excited Scalar, Pseudoscalar, and Vector Meson Nonets in a Chiral Quark Model // Fiz. Elem. Chast. At. Yadra. 2000. V. 31. No. 3. P. 576-633.

\bibitem{ven} \textit{Volkov M.K., Ebert D., Nagy M.} Excited Pions, $\rho-$ and $\omega-$ Mesons and Their Decays in a Chiral $SU(2)\times SU(2)$ Lagrangian // Int. J. Mod. Phys. A. 1998. V. 13. P. 5443. 

\bibitem{akv1} \textit{Arbuzov A.B.,Kuraev E.A.,Volkov M.K.} Processes        $e^+ e^- \rightarrow \pi^0 ({\pi^0}')\gamma$ in the NJL Model // Eur. Phys. J. A. 2011. V. 47. P. 103.

\bibitem{akv2} \textit{Arbuzov A.B.,Kuraev E.A.,Volkov M.K.} Production of $\omega \pi^0$ Pair in Electron-Positron Annihilation //  Phys.Rev. C. 2011. V. 83. P. 048201.

\bibitem{arb} \textit{Arbuzov A.B., Volkov M.K.} Two-Photon Decays and Photoproduction on Electrons of $\eta(550), \eta'(958), \eta(1295), \eta(1475)$ mesons // Phys. Rev. C. 2011. V. 84. P. 058201.

\bibitem{eepipi} \textit{Volkov M.K., Kostunin D.G.} The Processes $e^+ e^- \rightarrow \pi\pi(\pi')$ in the Extended NJL Model // Phys. Rev. C. 2012. V. 86. P. 025502.

\bibitem{taupipi} \textit{Volkov M.K., Kostunin D.G.} $\tau^- \rightarrow\pi^- \pi^0 \nu_{\tau}$ Decay in the Extended NJL Model // Fiz. Elem. Chast. At. Yadra, Lett. 2013. V.10. No. 1(178). P. 18-23.

\bibitem{eta} \textit{Ahmadov A.I., Kostunin D.G., Volkov M.K.} Processes of $e^+ e^- \rightarrow \eta, \eta', \eta(1295), \eta(1475))\gamma$ in the Extended Nambu--Jona-Lasonio Model // Phys. Rev. C. 2013. V. 87. P. 045203.

\bibitem{matrix} \textit{Volkov M.K., Yudichev V.L.} Radial Excitations of Scalar and $\eta, \eta'$ Mesons in a Chiral Quark Model // Yad. Phys. 2000. V.63. No. 10. P. 1924-1935.

\bibitem{pdg} \textit{Beringer J. et al.} Particle Physics Booklet // Phys. Rev. D. 2012. V. 86. P. 010001.

\end{thebibliography}
\end{document}